\documentclass[referee,useAMS,usenatbib,usegraphicx]{mn2e}
\usepackage{}
\usepackage{amsmath}
\usepackage{amssymb}
\usepackage{cases}

\title[Inclination-angle of the outflow in IRAS 05553+1631: A method to correct the projection effect]{Inclination-angle of the outflow in IRAS 05553+1631: A method to correct the projection effect}

\author[Shuo Kong, Yuefang Wu]{Shuo Kong$^{1}$\thanks{E-mail: protosswon@gmail.com}, Yuefang Wu$^{2}$\thanks{E-mail: ywu@pku.edu.cn}\\
$^{1}$School of Physics, Peking University, Beijing 100871, China. \\
$^{2}$Department of Astronomy, School of Physics, Peking University, Beijing 100871, China. }
\begin{document}

\date{Accepted by the Monthly Notices of the Royal Astronomical Society}

\pagerange{\pageref{firstpage}--\pageref{lastpage}} \pubyear{2002}

\maketitle

\label{firstpage}

\begin{abstract}
A mapping study of IRAS 05553+1631 was performed with $^{12}$CO J=3-2 and $^{13}$CO J=2-1 lines observed by the KOSMA 3 m-telescope. A core with a size of 0.65 pc and with a LTE mass of 120 M$_\odot$ was defined by the mapping with $^{13}$CO J=2-1 line. We have identified a bipolar outflow with $^{12}$CO J=3-2. For accuracy in the calculation of outflow parameters, overcoming the projection effect is important. We propose a new method to directly calculate the inclination-angle $\theta$. We establish two basic equations with the help of outflow contour diagram and finally obtain the "angle function" and the "angle equation" to derive $\theta$. We apply our method to the outflow of IRAS 05553+1631, finding that $\theta_{blue}$ is 73$^\circ$ and $\theta_{red}$ is 78$^\circ$. Compared to the parameters initially estimated under an assumption of 45$^\circ$ inclination-angle, the newly derived parameters are changed with different factors. For instance, the timescales for the blue and the red lobes are reduced by 0.31 and 0.21, respectively. Larger influences apply to mechanical luminosity, driving force, and mass-loss rate. The comparisons between parameters before and after the correction show that the effect of the inclination-angle cannot be neglected.
\end{abstract}

\begin{keywords}
stars: formation -- ISM: jets and outflows.
\end{keywords}

\section{Introduction}

Massive star formation (MSF) has attracted much attention. It has enormous impact on the natal interstellar medium (ISM) and on the evolution of Galaxy. MSF is believed to originate in dense molecular cores (DMCs) that can be traced by CO and its isotopologues. With the help of mapping, we can reveal the structure of molecular cores and investigate the MSF activity taking place within them.

High velocity outflows have also been intensively studied. First uncovered in 1976 \citep{b19,b8}, molecular outflows toward massive young stellar objects (YSOs) have attracted much attention \citep{b16}. Perhaps tracing the earliest stage \citep{b9} of star formation, molecular outflows are critical in the debate about two mechanisms for MSF: massive stars forming through accretion-disk-outflow or via collision-coalescence \citep{b22,b23}. The properties of outflows reveal the mass-loss phase before the main sequence. What we observe in the sky is not the real outflow but its two-dimensional projection. Therefore, it is essential to know the inclination-angle $\theta$ between the outflow axis and our line-of-sight. For instance, it will introduce a factor of 1/cos$\theta$ to the momentum. So far the determination of the inclination-angle has been somewhat neglected though some previous researchers have made a few attempts by modeling \citep{b3,b10}.

Until now, authors usually took assumptions for the inclination-angle. For instance, \citet{b7} hypothesized 45$^\circ$ while \citet{b6} supposed 60$^\circ$. The outflow parameters derived from these assumptions can only be meaningful statistically, but they are not accurate. In addition, work has been done to study the collimation of outflows \citep{b1,b15}. However, the projection effect should also be included.

In order to directly calculate the inclination-angle and to eliminate or reduce outflow parameter uncertainties, we propose a new method in this paper. In the next section, we will describe our observations of IRAS 05553+1631. In section \ref{sec:results}, the results will be presented. We will introduce our method and discuss some of its properties in section \ref{sec:discussion}. Section \ref{sec:summary} is a brief summary of our work.

\section{Observation}

The observations were performed with the KOSMA 3 m submillimeter telescope at the Gornergrat Observatory in Switzerland in 2002 to 2004. The dual-channel-SIS receiver was used. We observed simultaneously transitions in the 230 GHz tuning range ($^{13}$CO J=2-1) and in the 345 GHz tuning range ($^{12}$CO J=3-2). The beam sizes were 120" at 230 GHz and 80" at 345 GHz. The beam efficiencies for 230 GHz is 0.68 and for 345GHz is 0.72. The integration time for each point was about 1.5 min for both $^{12}$CO J=3-2 and $^{13}$CO J=2-1. The system temperatures were about 160 K for $^{13}$CO J=2-1 and 235 K for $^{12}$CO J=3-2. The spectral resolutions were 0.22 km/s and 0.29 km/s for $^{13}$CO J=2-1 and $^{12}$CO J=3-2, respectively. We used the position-switch mode in the observations. The map step size is 1 arcmin.

\section{Results}\label{sec:results}

\begin{figure*}
\centering
\includegraphics[width=150mm]{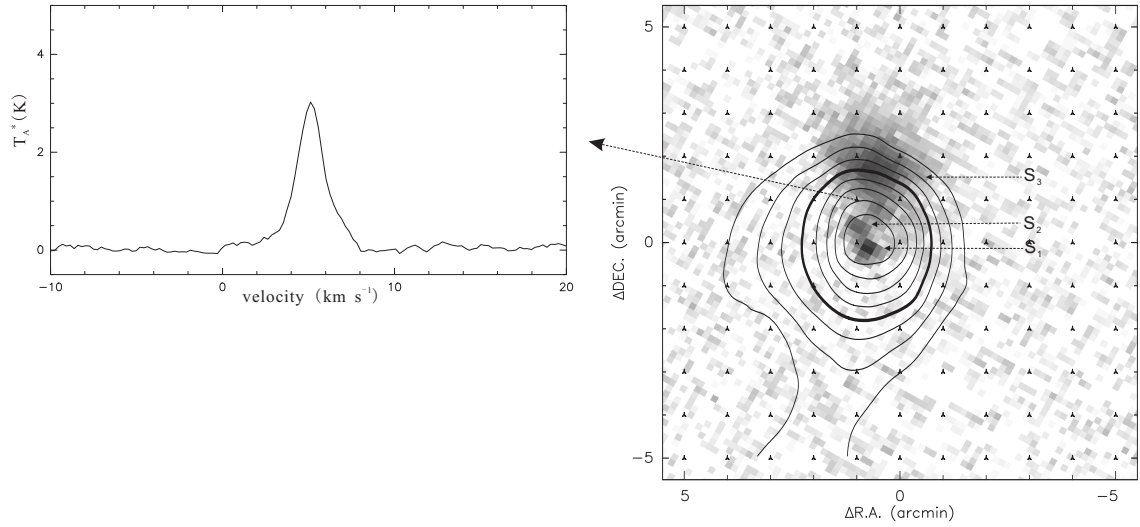}
\caption{$^{13}$CO J=2-1 spectrum and contour diagram. The contours range from 20 per cent to 90 per cent of the peak integration in steps of 10 per cent. The thick line shows the half intensity contour. The grey-scale background is the MSX 8.28 $\mu m$ emission. The three MSX sources are denoted S$_1$, S$_2$, and S$_3$, respectively.}\label{fig:13CO}
\end{figure*}

Both $^{13}$CO J=2-1 and $^{12}$CO J=3-2 emission were detected and mapped. The typical $^{13}$CO spectrum is shown in the left panel of Figure \ref{fig:13CO}. There is a single component in this molecular region. Table \ref{tab:observed} summarizes the parameters. Column (1) is the IRAS source name. Columns (2) to (5) are the positions of the source, including both the 1950 coordinates and the 2000 coordinates. Columns (6) to (8) list the $^{13}$CO J=2-1 spectrum parameters derived from Gaussian fitting. Columns (9) to (11) exhibit the IRAS flux parameters. According to \citet{b14}, IRAS sources that satisfy Log(F$_{25}$/F$_{12}$)$>$0.57, Log($F_{60}$/F$_{12}$)$>$1.30 and that peak at 100 $\mu m$ are promising candidates for UC H\,{\scriptsize II} regions. From Table \ref{tab:observed} we can see that the first two criteria are satisfied. We confirm that IRAS 05553+1631 has its maximum flux at 100 $\mu m$. The results show that IRAS 05553+1631 is a candidate for UC H\,{\scriptsize II} region. The right panel of figure \ref{fig:13CO} shows the core contours. The grey-scale background is MSX 8.28 $\mu m$ emission. The three MSX sources are named as S$_1$, S$_2$, and S$_3$, respectively. Spatially, S$_1$ and S$_2$ are both possibly associated with the core. We also made use of the Two Micron All Sky Survey (2MASS) data. S$_1$ has one 2MASS counterpart 05581473+1631070 whose magnitudes in the J, H, and K$_s$ bands are 13.103, 10.752, and 9.041, respectively; S$_2$ has one 2MASS counterpart 05581574+1631373 whose magnitudes in the J, H, and K$_s$ bands are 12.571, 11.886, and 11.565, respectively. S$_1$ is the reddest source and more likely to be correlated with the molecular core.

\begin{figure*}
\centering
\includegraphics[width=160mm]{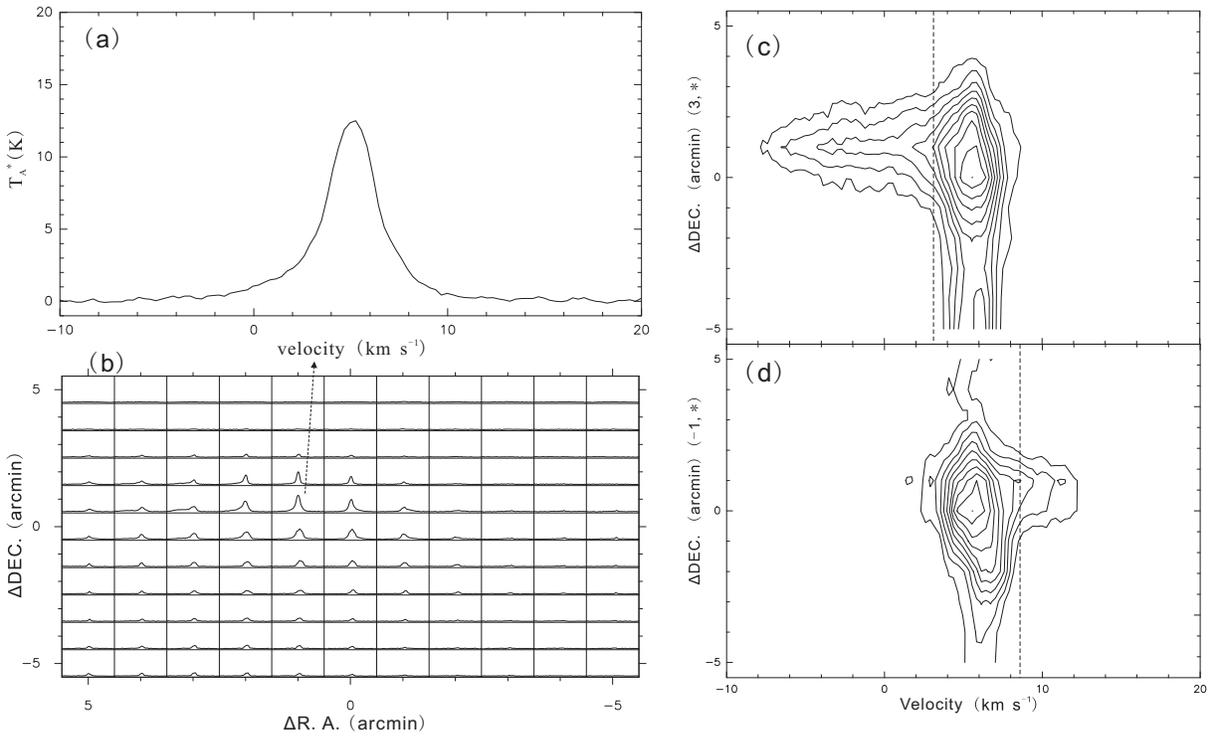}
\caption{$^{12}$CO J=3-2 results for IRAS 05553+1631. \textbf{(a)} spectrum. \textbf{(b)} $^{12}$CO J=3-2 grid map. \textbf{(c)(d)} $^{12}$CO P-V diagrams. They are along two different longitudes indicated beside the position axes. The dashed lines indicate the start of the blue and red lobes, respectively.}\label{fig:12CO}
\end{figure*}

\begin{figure}
\centering
\includegraphics[width=80mm]{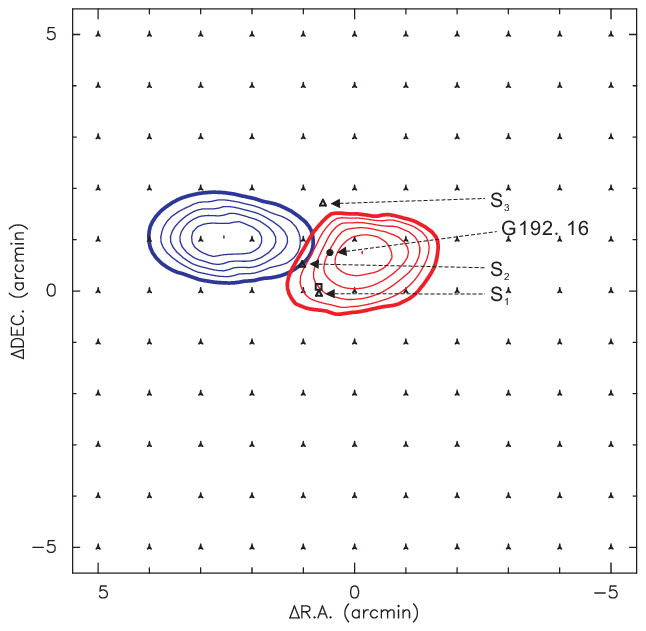}
\caption{Outflow contour diagram. The east blue contours represent the blue outflow lobe while the west red contours represent the red lobe. The peaks for the blue and red lobe are 14.6 K km s$^{-1}$ and 4.3 K km s$^{-1}$. The triangles show the positions of the three MSX sources. The square stands for the position of the $^{13}$CO core centre. The black solid circle presents the position of G192.16.}\label{fig:outflow}
\end{figure}

Figure \ref{fig:12CO} presents the $^{12}$CO J=3-2 results. In the P-V diagrams (c) and (d), the wide wings are obvious. The velocity ranges of the high-velocity gas were determined as -8 km s$^{-1}$ to 3.1 km s$^{-1}$ and 8.5 km s$^{-1}$ to 13 km s$^{-1}$ (hereafter refered to as HV-ranges) for blue and red lobes, respectively. The HV-ranges were integrated into the outflow contour diagram (Figure \ref{fig:outflow}). The outermost blue and red contours are 50 per cent of the blue and red peak intensities, respectively. The contours show a bipolar structure. \citet{b13} has identified YSO G192.16 as the driving source. Both their work (see Figure 1 of their paper) and our result (Figure \ref{fig:outflow}) show that the blue lobe is more extended from G192.16 than the red one. G192.16 is in good alignment between the two outflow lobes, and we adopt G192.16 as the driving centre.

The core parameters are summarized in Table \ref{tab:core}. Column (2) lists the 1.4 kpc distance calculated using the Galactic rotation model \citep{b2} while \citet{b13} gave a distance of 2.0 kpc. Column (3) gives the velocity range for the $^{13}$CO J=2-1 line. Column (4) displays the core radius which is half the mean value of two linear lengths of the half maximum contour. Columns (5) to (8) are derived from the radiation transfer equation. A ratio of 8.9$\times$10$^5$ for [H$_2$/$^{13}$CO] was adopted to calculate the H$_2$ column density (column 8). In column (9), we give the H$_2$ density calculated from the column density and the diameter (which is twice the radius). In the calculation of LTE mass which is shown in column (10), the mean atomic weight factor of 1.36 was introduced \citep{b6}. In column (11), the virial mass was calculated following \citet{b5}.

Table \ref{tab:outflow} summarizes the outflow parameters. Column (1) indicates the outflow lobes. Column (2) lists the outflow sizes. The thick contour was covered by an ellipse whose major axis is the linear length from west to east and whose minor axis is the linear length from north to south. The outflow sizes are the lengths of semimajor axes. Column (3) lists the masses for blueshifted and redshifted gas, respectively. The beam averaged column density for the $^{12}$CO J=3-2 transition is
\begin{equation}
   \overline N=4.8\times10^{12}\frac{(T_{ex}+0.92)}{exp(-33.2/T_{ex})}\int\frac{T_a^*}{\eta_b}\frac{\tau}{[1-exp(-\tau)]}\mathrm{d}v
\end{equation}
where T$_{ex}$ is the core excitation temperature. The integration is over the HV-range. The $^{12}$CO J=3-2 transition is usually optically thick even in the line wings \citep{b12}, therefore we adopted a mean optical depth $\overline{\tau}$ of 4 \citep{b6}. For the mass calculation, an [H$_2$/CO] ratio of 10$^4$ was taken. Momentum and energy are displayed in columns (4) and (5). The adopted velocity $v$ is the line-of-sight velocity divided by cos$\theta$ where $\theta$ is the inclination-angle. Column (6) shows the dynamical timescales which were estimated as R/$v$ where R is the distance from the driving centre to the peak of high-velocity gas. The mechanical luminosity, driving force, and mass-loss rate are listed in columns (7) to (9), respectively. They were calculated as E/t, P/t, and P/(tV$_w$), respectively \citep{b17}, where V$_w$ is the wind velocity assumed to be 100 km/s. In the calculation, we first assumed $\theta$=45$^\circ$, the calculation will be improved with our method (see section \ref{sec:outflow}).

\section{Discussion}\label{sec:discussion}

\subsection{Core}

Table \ref{tab:core} shows that the core has a size $\sim$0.65 pc and that M$_{LTE}$ is equal to 120 $M_{\odot}$. The $^{13}$CO line width is 2.5 km s$^{-1}$ which is larger than those of low-mass cores \citep{b11}. These results suggest that the core is massive. According to virial theory, a core is gravitationally bound when its mass is larger than its virial mass. In our results, this is not satisfied, which is possibly due to the [H$_2$/$^{13}$CO] ratio adopted or the optically thick assumption for $^{13}$CO J=2-1 emission; alternatively, the core may be unbound.

\subsection{Outflow}\label{sec:outflow}

In our calculation for outflow parameters, 45$^\circ$ was assumed as the inclination-angle. Actually, almost all the outflow parameters depend on the angle. For instance, the velocity of outflow gas is the projection along our line-of-sight. Therefore a factor of 1/cos$\theta$ should be introduced. The same goes for momentum, timescale, etc., but sometimes with different factors. The inclination-angle also affects the collimation factor. Thus, accurate determination of $\theta$ is important. Unfortunately, until now, people assume a value for $\theta$ \citep{b7,b6,b18}. In the following, we propose a new method to improve this situation.

\subsubsection{The inclination-angle $\theta$}

Our method to derive $\theta$ requires knowledge of the outflow geometry, so before the calculation, it is necessary to discuss something about the shapes of outflows. Generally, without impact from the surrounding materials, the molecular outflows are likely to be axially symmetric. This could be inferred from the symmetry of the outflow contour diagrams (\citet{b7}, Figure 2; \citet{b17}, Figure 5c; \citet{b1}, Figure 13 NGC 2071) and from modeling \citep{b20,b10,b21}. In our results, the P-V diagram (Figure \ref{fig:12CO}) and the outflow contours (Figure \ref{fig:outflow}) show good symmetry, especially the blue lobe. In addition, the ellipse-like outflow contours indicate a projection of a conical outflow (see Section \ref{sec:more}). Since a cone is an ideal form for an axially symmetric outflow (see Section \ref{sec:comparison}, Paragraph 2), we assume that the shape of the outflow is a cone. We also assume that high velocity components form cone-shell like structure whose emission intensity decreases from inside to outside. The cone could be called a "cone onion".

\begin{figure*}
\centering
\includegraphics[width=130mm]{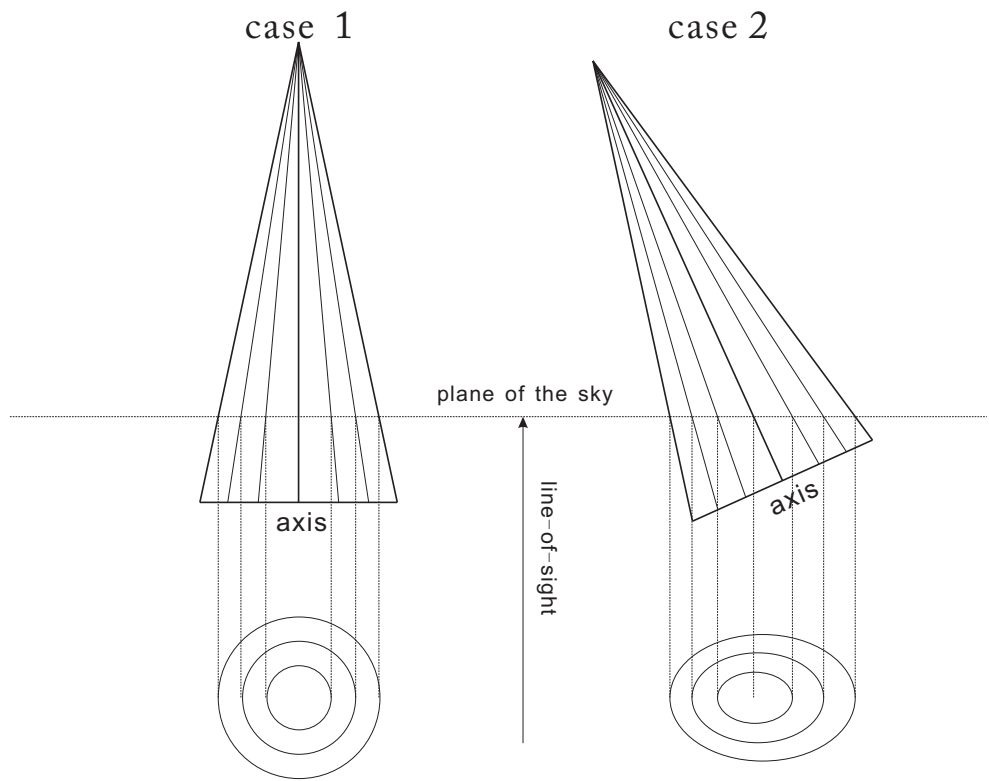}
\caption{Illustration of the geometry. The cones represent outflows with different cases. Observer is looking up from the bottom of the figure. The circles and ellipses are what the observer sees in the plane of the sky in the two cases. Case 1 represents the outflow axis aligned with the line-of-sight. In case 2, the outflow axis is inclined. It is obvious that in case 2, the ellipses are not co-centred. And the projection point of axis is not at the centre of the ellipse.}\label{fig:cone}
\end{figure*}

\begin{figure*}
\centering
\includegraphics[width=120mm]{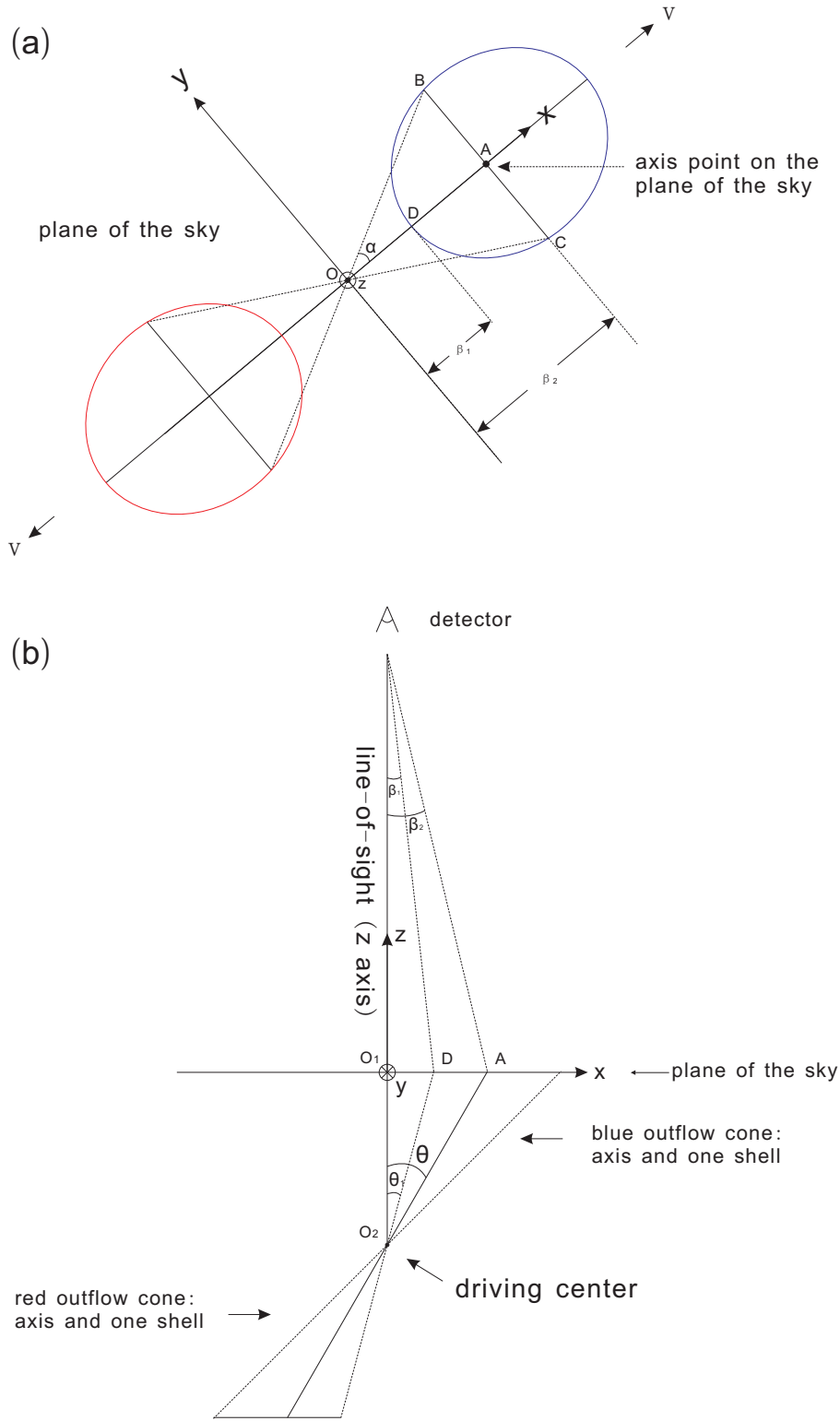}
\caption{Illustrations for the calculation. \textbf{(a)} Outflow contours on the plane of the sky. The blue ellipse on the right represents the contour of the blueshifted gas while the red ellipse on the left represents the contour of the redshifted gas. Point O shows the position of the driving centre. The xyz-coordinate system is centred on the driving centre. \textbf{(b)} The projection plane perpendicular to the y-axis. $\theta$ is the inclination-angle. Points A and D in \textbf{(b)} are the same as those in \textbf{(a)}. $\bigodot$ represents axis pointing outward perpendicular to the paper while $\bigotimes$ represents axis pointing inward.}\label{fig:model}
\end{figure*}

When we observe, what we see is the outflow projection on the sky. Thus, if the outflow axis is at an angle $\theta$ to our line-of-sight, the contours will appear as ellipses (Figure \ref{fig:cone}; the contours could also be parabola or hyperbola, see section \ref{sec:more}). Figure \ref{fig:model} presents the calculation of the inclination-angle $\theta$. In Figure \ref{fig:model}(a), one can see the driving centre and an elliptical outflow contour. From Figure \ref{fig:model}(b), a relation of the four angles: $\beta_1$, $\beta_2$, $\theta_1$, and $\theta$ can be expressed by equation (\ref{equ:4angle})
\begin{equation}\label{equ:4angle}
   \frac{\tan\theta_1}{\tan\theta}=\frac{\tan\beta_1}{\tan\beta_2}\approx\frac{\beta_1}{\beta_2}
\end{equation}
where $\beta_1$ and $\beta_2$ are the angle distances in the contour-diagram ($\beta_1$ accounts for the angle between driving centre and point D; $\beta_2$ is the angle between driving centre and axis point A). $\beta_1$ and $\beta_2$ are usually very small (about several arcmin), thus the small angle approximation is valid. $\theta_1$ and $\theta$ are unknown. Then, we have to find another equation for either $\theta_1$ or $\theta$. Notice that the length of line AB does not change as a function of $\theta$. In Figure \ref{fig:model}(a), we have
\begin{equation}\label{equ:a}
   \overline{AB}=\overline{AO}\tan\alpha
\end{equation}
where $\alpha$ is the angle between OA and OB. Overlines indicate length. If the outflow contour is not symmetric about line AO, we take $\alpha$ as half the angle $\angle BOC$. We have already assumed the shape of cone, so the opening angle $\theta$-$\theta_1$ is the same for the entire cone. Therefore in Figure \ref{fig:model}(b) we have
\begin{equation}\label{equ:b}
   \overline{AO_1}=\overline{AO_2}\sin\theta
\end{equation}
\begin{equation}\label{equ:c}
   \overline{AB}=\overline{AO_2}\tan(\theta-\theta_1)
\end{equation}
where O$_1$ and O$_2$ are the same points as O in the plane of the sky. $\overline{AO_1}$ is equal to $\overline{AO}$. Thus, combining equations (\ref{equ:a}), (\ref{equ:b}), (\ref{equ:c}), and substitute $\overline{AO_1}$ with $\overline{AO}$ we have
\begin{equation}\label{equ:alpha}
   \tan\alpha=\frac{\tan(\theta-\theta_1)}{\sin\theta}
\end{equation}
Putting together equation (\ref{equ:4angle}) and equation (\ref{equ:alpha}), we can solve the inclination-angle $\theta$ by eliminating $\theta_1$. Since $\beta_1$, $\beta_2$, and $\alpha$ can be derived from the outflow contour, we try to further manipulate the two equations to obtain a better expression. We define the "angle constants"
\begin{equation}
   P=\frac{\beta_1}{\beta_2}
\end{equation}
and
\begin{equation}
   Q=\tan\alpha
\end{equation}
Combining equations (\ref{equ:4angle}) and (\ref{equ:alpha}) and eliminating $\theta_1$, we have
\begin{equation}
   \frac{1-P}{Q}\frac{1}{\cos\theta}-P\tan\theta\tan\theta=1
\end{equation}
Defining the "angle function"
\begin{equation}\label{angle function}
A(\theta)=\frac{1-P}{Q}\frac{1}{\cos\theta}-P\tan\theta\tan\theta,
\end{equation}
after determining P and Q, we obtain the inclination-angle by solving the "angle equation"
\begin{equation}\label{angle equation}
A(\theta)=1
\end{equation}

\subsubsection{Comparison}\label{sec:comparison}

\begin{figure*}
\centering
\includegraphics[width=100mm]{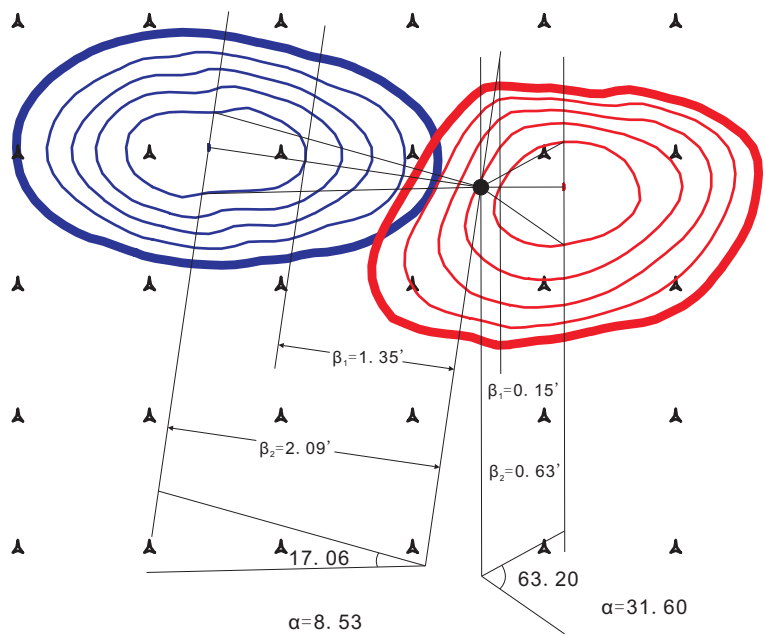}
\caption{Calculation of the inclination-angle. The blue contours on the left represent the blue lobe while the red ones on the right represent the red lobe. The black solid circle is the driving centre G192.16. The relevant angles are presented in the figure.}\label{fig:theta}
\end{figure*}

We utilize our method on IRAS 05553+1631. Figure \ref{fig:theta} shows the derivation process. In general, the shapes and positions in Figure \ref{fig:theta} are similar to those in our model (Figure \ref{fig:model}(a)). Table \ref{tab:comparison} presents the comparison between newly derived and old outflow parameters. We find the inclination-angles to be 73$^\circ$ and 78$^\circ$ for blue and red lobes, respectively. This suggests that the two outflow lobes are in good alignment. In Table \ref{tab:comparison}, compared with the former results, the newly derived parameters have rather large corrections. For the blue lobe, the momentum is enlarged by a factor of (cos45$^\circ$/cos73$^\circ\approx$) 2.4, as the velocity is corrected by 1/cos$\theta$. The correction for energy is proportional to that of P$^2$ which is about 5.8. The timescale is reduced by a factor of (cot73$^\circ$/cot45$^\circ\approx$) 0.31. The correction factors for mechanical luminosity, driving force, and mass-loss rate are 19, 7.7, and 7.7, respectively. For the red lobe, the analysis is similar, but the factors are different. These results suggest that the correction from inclination-angle cannot be neglected and our method is practical.

From the process of our method one can see that the cone assumption ensure the viability of equation (\ref{equ:alpha}). If not, the equation would be an approximation. Let $\theta_0$=$\theta$-$\theta_1$, the Taylor expansion of tan$\theta$ around $\theta_0$ is (keeping the first order) tan$\theta$=tan$\theta_0$+$\frac{\Delta\theta}{(cos\theta_0)^2}$. For the blue outflow lobe in our case, tan$\theta_0$=0.143, if $\Delta\theta$=1$^\circ$ (0.017 radians), then the second term of the Taylor expansion would be $\sim$0.018, the uncertainty is about 12 per cent. For the red outflow lobe, the uncertainty is about 4 per cent.

\subsubsection{More properties}\label{sec:more}

Geometrically, when a cone is cut by a plane, the resulting curve can be an ellipse (including a circle), a parabola, or a hyperbola, depending on the angle between the cone-axis and the plane. For an ellipse, the larger the inclination-angle is, the larger its eccentricity will be, and vice versa. When $\theta$ approaches 90$^\circ$ we obtain a parabola or a hyperbola. Theoretically, all of the curves can be utilized for the calculation. But the ellipse has the advantage that it is usually much more conspicuous and easy to manipulate. The parabola and the hyperbola, on the contrary, can be confusing because of the large inclination-angle. The outflow edge from the driving source has very weak emission and the contours exhibit a fan-like structure. The centre and the curve will be ambiguous. The opposite extreme is when the inclination-angle becomes zero and the ellipse changes to a circle. Fortunately, the angle equation (\ref{angle equation}) still works. In the angle function (\ref{angle function}), the coefficient of the first term $\frac{1-P}{Q}$ is ($\frac{1-\frac{\overline{OD}}{\overline{OA}}}{\frac{\overline{AB}}{\overline{OA}}}$=$\frac{\overline{OA}-\overline{OD}}{\overline{AB}}$=$\frac{\overline{AD}}{\overline{AB}}$). As $\theta$ approaches zero, the projection shape becomes a circle and $\frac{\overline{AD}}{\overline{AB}}\rightarrow$1. Meanwhile, cos$\theta$ $\rightarrow$1 and the second term vanishes. Thus A($\theta\rightarrow$0)$\rightarrow$1. The angle equation (\ref{angle equation}) is tenable in this extreme case. However, when $\theta$ approaches zero, use of this method could be problematic if the resolution is low. Another property is the collimation. In our view, it can be manifested by the opening angle $\theta$-$\theta_1$. Larger $\theta$-$\theta_1$ means lower collimation, and vice versa. Thus it is reasonable to define a collimation factor as cot($\theta$-$\theta_1$) (=(tan$\alpha$sin$\theta$)$^{-1}$). In IRAS 05553+1631, the factors are 7.0 and 1.6 for blue and red lobes, respectively. One thing should be mentioned, the ellipses we used for IRAS 05553+1631 are 90 per cent contours (Figure \ref{fig:theta}) as stronger emission could reduce the errors. If we chose the 50 per cent contours, the factor would change. Additionally, the method highly depends on the accuracy with which the driving centre is located. One question is the identification, and the second is the spatial resolution. The identification involves personal judgement, which is accompanied by considerable uncertainty in some cases. \citet{b4} showed that 6.7 GHz methanol maser and relevant 24 $\mu m$ emission usually coincide with each other and they are good tracers to the driving centre, this may offer much help for the identification. The resolution is usually low for single-dish observations. Using high-resolution telescopes and interferometers can improve this.

\begin{figure}
\centering
\includegraphics[width=80mm]{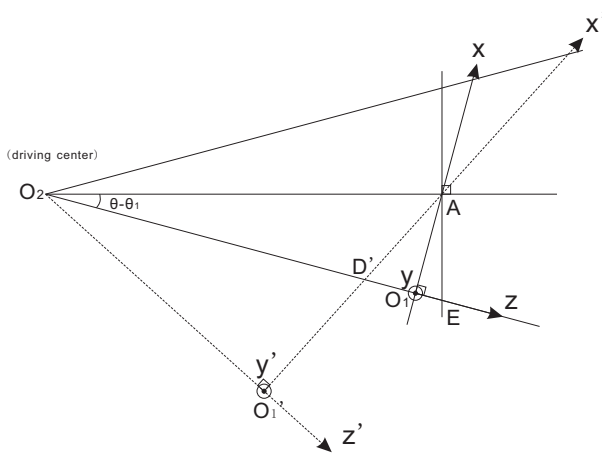}
\caption{The illustration of parameter Q. The utilization of letters follows Figure \ref{fig:model}. Condition c2 is denoted by prime.}\label{fig:Q}
\end{figure}

\begin{figure*}
\centering
\includegraphics[width=160mm]{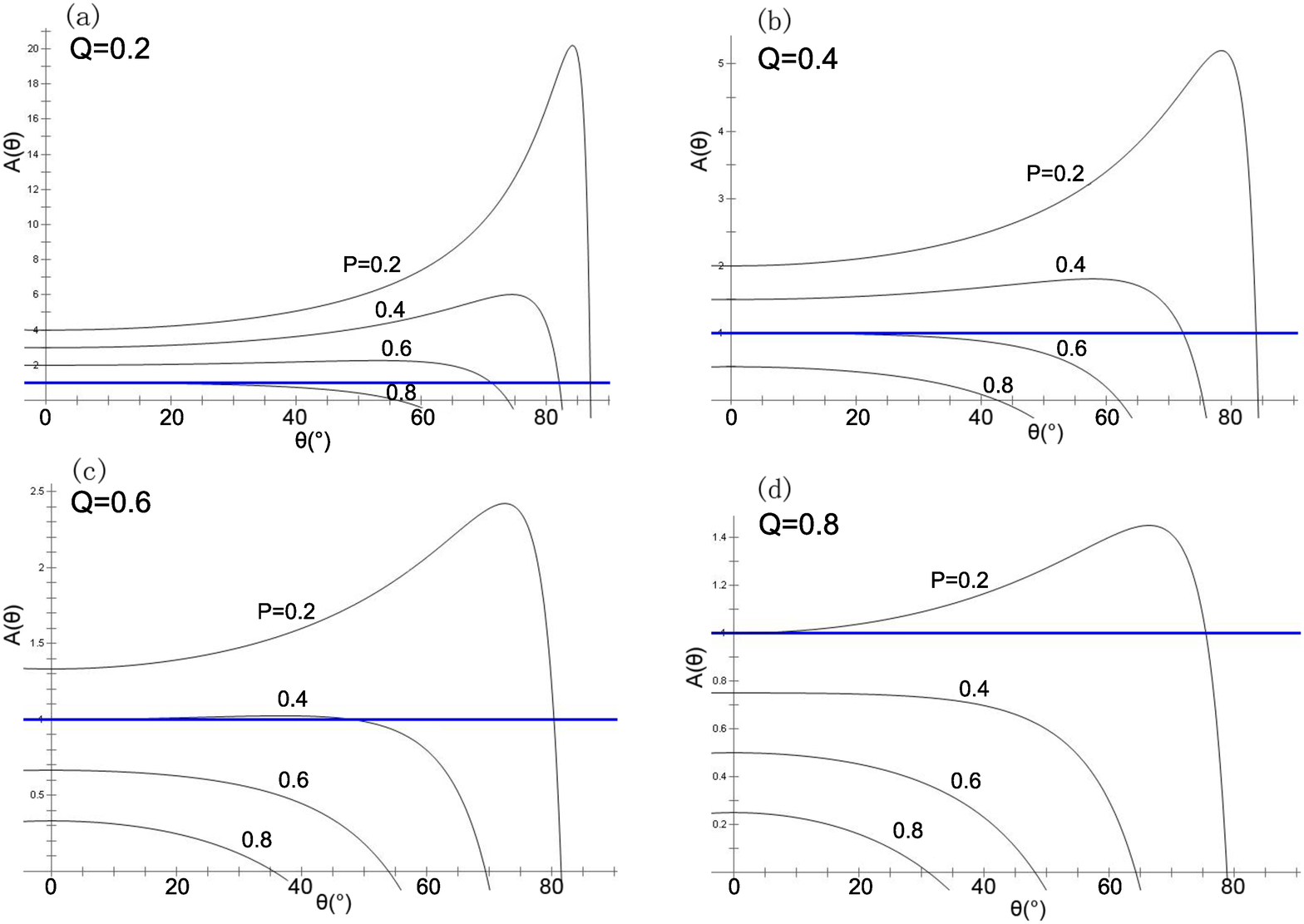}
\caption{Angle function curves with different P and Q. Each of \textbf{(a)(b)(c)(d)} shows the variation of the angle function with the changing of P (from 0.2 to 0.8 with an interval of 0.2) and with a fixed Q ( \textbf{(a)} Q=0.2; \textbf{(b)} Q=0.4; \textbf{(c)} Q=0.6; \textbf{(d)} Q=0.8). The $\theta$ axis is in angle system. The blue straight line shows A($\theta$)=1. The $\theta$ of the intersection point indicates the solution to the angle equation.}\label{fig:curves}
\end{figure*}

In the angle function (\ref{angle function}), there are two parameters P and Q. Since we can always choose an ellipse excluding the driving centre, it is reasonable to assume 0$<$P$<$1. Figure \ref{fig:Q} illustrates the value of Q, which shows the projection on x-z plane. Lines AO$_1$ and AD' represent the projections of two conditions for the plane of the sky: (1) AO$_1$ perpendicular to O$_2$E (solid, hereafter named c1); (2) AD' not perpendicular to O$_2$E (dashed, hereafter named c2). Condition c2 is denoted by prime. $\overline{AE}$ equals the $\overline{AB}$ of Figure \ref{fig:model}(a). Thus, it is easy to see that the maximum of Q is $\frac{\overline{AE}}{\overline{AO_1}}$ which is just 1/cos($\theta$-$\theta_1$). Since ($\theta$-$\theta_1$) cannot be large and an angle of 60$^\circ$ can only bring Q=2, we consider 0$<$Q$<$2. When P=0, the second term of the angle function (\ref{angle function}) vanishes and $\theta$=arccos(1/Q). In order to have solution, Q must be larger than 1. In fact, in this case Q reaches its maximum (see Figure \ref{fig:Q}). The result is 0$<\theta<$71$^\circ$. When P=1, there is no solution. Figure \ref{fig:curves} shows the variation of the angle function as a function of P and Q. With a fixed Q, the function decreases as P increases. With a fixed P, the function decreases when Q increases. Usually within (0$^\circ$,90$^\circ$), there is one solution. When Q approaches 1 (Figure \ref{fig:curves}(d)), in order to have solution, P must be closer to 0. When P approaches 1, the same happens to Q. The variation of angle function is sensitive to the values of P and Q.

\section{Summary}\label{sec:summary}

We mapped IRAS 05553+1631 with $^{12}$CO J=3-2 and $^{13}$CO J=2-1 lines. A core was identified from $^{13}$CO J=2-1 observations. It has a size of 0.65 pc and LTE mass of 120 M$_{\odot}$ which is lower than the virial mass of 850 M$_{\odot}$. $^{12}$CO J=3-2 mapping revealed a bipolar outflow. Its parameters were initially estimated under the assumption of a 45$^\circ$ inclination-angle. A new method to directly calculate the inclination-angle $\theta$ was proposed, and was utilized for the bipolar outflow of IRAS 05553+1631. We found that $\theta_{blue}$ is 73$^\circ$ and $\theta_{red}$ is 78$^\circ$. Parameters with the new $\theta$s were compared with the former ones. For the blue lobe, the momentum was enlarged from 82 M$_\odot$~km~s$^{-1}$ to 200 M$_\odot$~km~s$^{-1}$ by a factor of 2.4 while the timescale was reduced from 8.8$\times$10$^4$ yrs to 2.7$\times$10$^4$ yrs by a factor of 0.31. The enlarging factors for energy, mechanical luminosity, driving force, and mass-loss rate are 5.8, 19, 7.7, and 7.7, respectively. For the red lobe, the momentum was enlarged from 11 M$_\odot$~km~s$^{-1}$ to 36 M$_\odot$~km~s$^{-1}$ by a factor of 3.4 while the timescale was reduced from 6.3$\times$10$^4$ yrs to 1.3$\times$10$^4$ yrs by a factor of 0.21. The enlarging factors for energy, mechanical luminosity, driving force, and mass-loss rate are 12, 55, 16, and 16, respectively. The results show that a selection of parameters were influenced by the inclination-angle $\theta$.

\section*{Acknowledgments}

We are grateful for Dr. Martin Miller for the assistance of observations. We also thank Tie Liu, Zhiyuan Ren, and Xueying Tang for the helpful discussions. Thank Hongping Du for the language checking. Thank Prof. P. Goldsmith for the constructive suggestions and thank Bella Lock for the helpful work. This project is supported by grant 10733030 and 10873019 of NSFC.

\bsp

\newpage


\begin{table*}
\begin{minipage}{200mm}
\caption{Observed parameters.}\label{tab:observed}
\tabcolsep 0.5mm \scriptsize
\begin{tabular}{ccccccccccc}
\hline
\hline\\[0.2mm]
{\parbox[t]{16mm}{\centering Source\\~\\(1)}} &
{\parbox[t]{14mm}{\centering R.A.\footnote{Columns (2) to (5) present the position. The coordinate years are in parentheses.}\\(1950)\\(2)}}                      &
{\parbox[t]{14mm}{\centering DEC.\\(1950)\\(3)}}                      &
{\parbox[t]{14mm}{\centering R.A.\\(2000)\\(4)}}                      &
{\parbox[t]{14mm}{\centering DEC.\\(2000)\\(5)}}                      &
{\parbox[t]{10mm}{\centering $V_{lsr}$\footnote{System velocity.}\\km/s\\(6)}}                   &
{\parbox[t]{10mm}{\centering $T^*_{A13}$\footnote{Antenna temperature of $^{13}$CO J=2-1 spectrum.}\\K\\(7)}}     &
{\parbox[t]{10mm}{\centering $\Delta V_{13}$\footnote{Full Width at Half Maximum (FWHM).}\\km/s\\(8)}}                 &
{\parbox[t]{14mm}{\centering Log($\frac{F_{25}}{F_{12}}$)\footnote{Columns (9) to (11) list parameters related to IRAS flux density.}\\~\\(9)}}   &
{\parbox[t]{13mm}{\centering Log($\frac{F_{60}}{F_{12}}$)\\~\\(10)}}  &
{\parbox[t]{8mm}{\centering F$_{100}$\\Jy\\(11)}}                      \\
\hline\\[0.01mm]

05553+1631 &	05~55~18.0	&	16~31~00	&	05~58~11.5	&	16~31~14	&	5.5 &	3.1 &2.5 & 1.70 &	2.52 & 528 \\

\hline\\[0.02mm]
\end{tabular}
\end{minipage}
\end{table*}

~\\
~\\
~\\
~\\
~\\

\begin{table*}
\begin{minipage}{200mm}
\caption{Core parameters.}\label{tab:core}
\tabcolsep 1.0mm \scriptsize

\begin{tabular}{ccccccccccc}
\hline
\hline\\[0.2mm]
{\parbox[t]{18mm}{\centering Source\\~\\(1)}} &
{\parbox[t]{7mm}{\centering D\footnote{Distance.}\\kpc\\(2)}} &
{\parbox[t]{10mm}{\centering Range\footnote{Integrated velocity range in $^{13}$CO J=2-1 line.}\\km s$^{-1}$\\(3)}} &
{\parbox[t]{10mm}{\centering Radius\footnote{Half the mean value of two linear lengths of the thick contour.}\\pc\\(4)}} &
{\parbox[t]{10mm}{\centering $T_{ex}$\footnote{Excitation temperature.}\\K\\(5)}} &
{\parbox[t]{10mm}{\centering $\tau_{13}$\footnote{Optical depth of $^{13}$CO J=2-1 line.}\\~\\(6)}} &
{\parbox[t]{13mm}{\centering $N_{13}$\footnote{Column density of $^{13}$CO.}\\$10^{15}cm^{-2}$\\(7)}} &
{\parbox[t]{13mm}{\centering $N_{H_2}$\footnote{Column density of H$_2$.}\\$10^{21}cm^{-2}$\\(8)}} &
{\parbox[t]{13mm}{\centering $n_{H_2}$\footnote{Density of H$_2$.}\\10$^3$ $cm^{-3}$\\(9)}} &
{\parbox[t]{9mm}{\centering $M_{LTE}$\footnote{Mass estimated under LTE assumption.}\\$M_{\odot}$\\(10)}} &
{\parbox[t]{9mm}{\centering $M_{vir}$\footnote{Virial mass \citep{b5}.}\\$M_{\odot}$\\(11)}} \\

\hline\\[0.01mm]

05553+1631  &1.4  &(3.5,7.3) &0.65 &25.8  &0.22  &6.8 &6.0     &3.0 &120  &850\\

\hline\\[0.02mm]
\end{tabular}
\end{minipage}
\end{table*}

~\\
~\\
~\\
~\\
~\\

\begin{table*}
\begin{minipage}{200mm}
\caption{Outflow parameters.}\label{tab:outflow}
\tabcolsep 1.0mm \scriptsize

\begin{tabular}{ccc ccc ccc}
\hline
\hline\\[0.2mm]
{\parbox[t]{7mm}{\centering Lobe\\~\\(1)}} &
{\parbox[t]{7mm}{\centering Size\footnote{Half the linear horizontal length of the 50 per cent outflow contour.}\\pc\\(2)}} &
{\parbox[t]{8mm}{\centering M\footnote{Outflow mass in the unit of M$_\odot$.}\\M$_\odot$\\(3)}} &
{\parbox[t]{15mm}{\centering P\footnote{Momentum.}\\M$_\odot$~km~s$^{-1}$\\(4)}} &
{\parbox[t]{12mm}{\centering E\footnote{Energy.}\\10$^{45}$ ergs\\(5)}} &
{\parbox[t]{10mm}{\centering t\footnote{Dynamical timescale. The unit is year.}\\10$^4$ yrs\\(6)}} &
{\parbox[t]{13mm}{\centering L$_{mech}$\footnote{Mechanical luminosity.}\\10$^{-1}$ L$_\odot$\\(7)}} &
{\parbox[t]{28mm}{\centering F\footnote{Driving force.}\\10$^{-4}$ M$_\odot$~km~s$^{-1}$~yr$^{-1}$\\(8)}} &
{\parbox[t]{25mm}{\centering $\dot{M}$\footnote{Mass-loss rate.}\\10$^{-6}$ M$_\odot$yr$^{-1}$\\(9)}} \\

\hline\\[0.01mm]

blue    &   0.65    &   6.2   &    82    &   11    & 8.8 &	 10    &   9.3	 &   9.3   \\
red     &   0.60    &   1.9   &    11    &~~~~0.60 & 6.3 &~~~~0.80 &   1.7   &   1.7   \\

\hline\\[0.02mm]
\end{tabular}
\end{minipage}
\end{table*}

~\\
~\\
~\\
~\\
~\\

\begin{table*}
\begin{minipage}{200mm}
\caption{Comparison between newly derived parameters and previous results for IRAS 05553+1631.}\label{tab:comparison}
\tabcolsep 1.0mm \scriptsize
\begin{tabular}{ccc ccc ccc c}
\hline
\hline\\[0.2mm]
{\parbox[t]{10mm}{\centering Lobe\\~\\(1)}} &                                         
{\parbox[t]{10mm}{\centering $\theta$\\$^\circ$\\(2)}} &                            
{\parbox[t]{7mm}{\centering Size\\pc\\(3)}} &                                         
{\parbox[t]{8mm}{\centering M\\M$_\odot$\\(4)}} &                                     
{\parbox[t]{15mm}{\centering P\\M$_\odot$~km~s$^{-1}$\\(5)}} &                     
{\parbox[t]{11mm}{\centering E\\10$^{45}$ ergs\\(6)}} &                               
{\parbox[t]{10mm}{\centering t\\10$^4$ yrs\\(7)}} &                                   
{\parbox[t]{13mm}{\centering L$_{mech}$\\10$^{-1}$ L$_\odot$\\(8)}} &                 
{\parbox[t]{28mm}{\centering F\\10$^{-4}$ M$_\odot$~km~s$^{-1}$~yr$^{-1}$\\(9)}} &    
{\parbox[t]{20mm}{\centering $\dot{M}$\\10$^{-6}$ M$_\odot$yr$^{-1}$\\(10)}} \\       

\hline\\[0.01mm]

new blue 	&	73	&	0.65	&   6.2   &  200~~   &	 64    & 2.7 &	190    &   74    &	 74    \\
new red 	&	78	&	0.60	&   1.9   &     36   &~~~~7.0  & 1.3 &~~~~43   &   27    &   27    \\
\hline\\[0.01mm]
old blue    & 45    &   0.65    &   6.2   &    82    &   11    & 8.8 &	 10    &   9.3	 &   9.3   \\
old red     & 45    &   0.60    &   1.9   &    11    &~~~~0.60 & 6.3 &~~~~0.80 &   1.7   &   1.7   \\

\hline\\[0.02mm]
\end{tabular}
\end{minipage}
\end{table*}


\label{lastpage}

\end{document}